\documentclass[runningheads]{svmult}

\usepackage{makeidx}   
\usepackage{graphicx}  
\usepackage{subeqnar}  
\usepackage{multicol}  
\usepackage{physprbb}  
\makeindex             

\newcommand{\Myr}{M$_{\odot}$\,yr$^{-1}$\ }         

\DeclareRobustCommand{\ion}[2]{
\relax\ifmmode                             
\ifx\testbx\f@series                       
{\mathbf{#1\,\mathsc{#2}}}\else            
{\mathrm{#1\,\mathsc{#2}}}\fi              %
\else\textup{#1\,{\mdseries\textsc{#2}}}
\fi}                                       

\begin{document}
\title*{The Multiband Photometry of GRB Host \protect\newline
Galaxies: Comparison with the Spectral \protect\newline
Energy Distributions of Nearby Galaxies \protect\newline
and Theoretical Modeling}

\toctitle{The Multiband Photometry of GRB Host \protect\newline
Galaxies: Comparison with the Spectral \protect\newline
Energy Distributions of Nearby Galaxies \protect\newline
and Theoretical Modeling}

\titlerunning{GRB Host Galaxies}

\author{V.~V.~Sokolov\inst{1,3}
\and T.~A.~Fatkhullin\inst{1}
\and V.~N.~Komarova\inst{1,3}
\and E.~R.~Kasimova\inst{2}
\and V.~I.~Korchagin\inst{2}}

\authorrunning{V. V. Sokolov et al.}

\institute{Special Astrophysical Observatory of RAS,
	   Karachai-Cherkessia, Nizhnij Arkhyz, 369167 Russia
      \and Institute of Physics, Rostov University,
	   Stachki 194, Rostov-on-Don, 344090, Russia
      \and Isaac Newton Institute of Chile, SAO Branch}

\maketitle

\begin{abstract}
We present one of the results of $BVRI$ photometry of the hosts of GRB  for
the host galaxy of GRB~970508
and the theoretical modeling of its continuum spectral energy distribution (SED)
to show that it is
important to take into account internal extinction in the host galaxies.
We compared the $BVRI$ broad-band flux spectrum of the host to
template SEDs of local starburst galaxies and found that there is
a significant internal extintion in this host. Moreover, this comparison
allows us to derive the absolute magnitude ($M_{B_{rest}}$) and rouhgly
estimate reddening ($A_V$). Population synthesis modeling of the continuum SED
for different reddening laws (\cite{Calz2000} and \cite{Cardelli})
demostrates that the observational data of the host galaxy of
GRB~970508 are best fitted by the spectral properties of a model SED with
extinction of $A_V\approx 2$.
\end{abstract}

\begin{figure*}[t]
\includegraphics[angle=-90,width=0.47\textwidth,bb=588 22 90 721,clip]{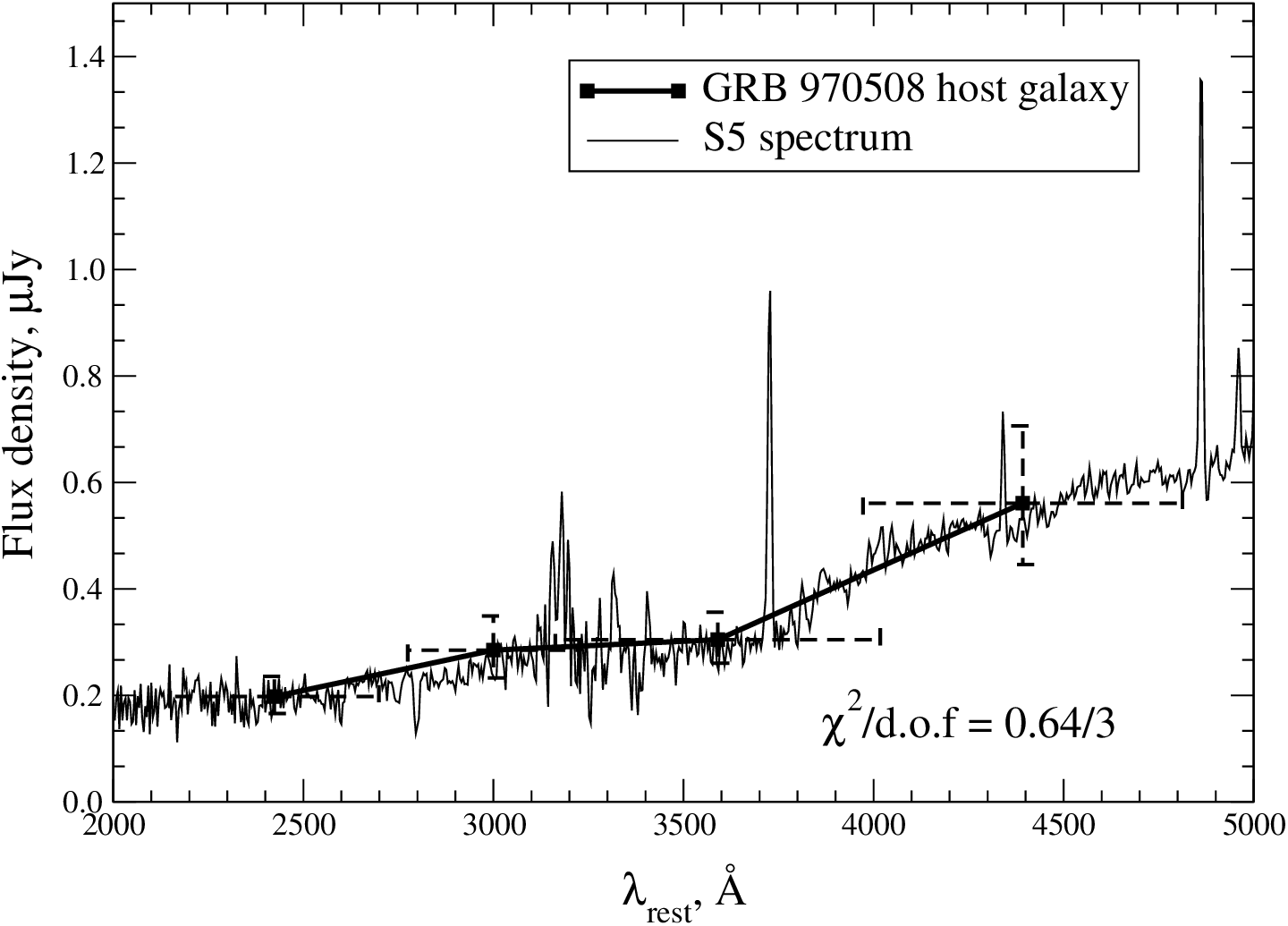}
\hfill
\includegraphics[width=0.47\textwidth,bb=30 25 722 530,clip]{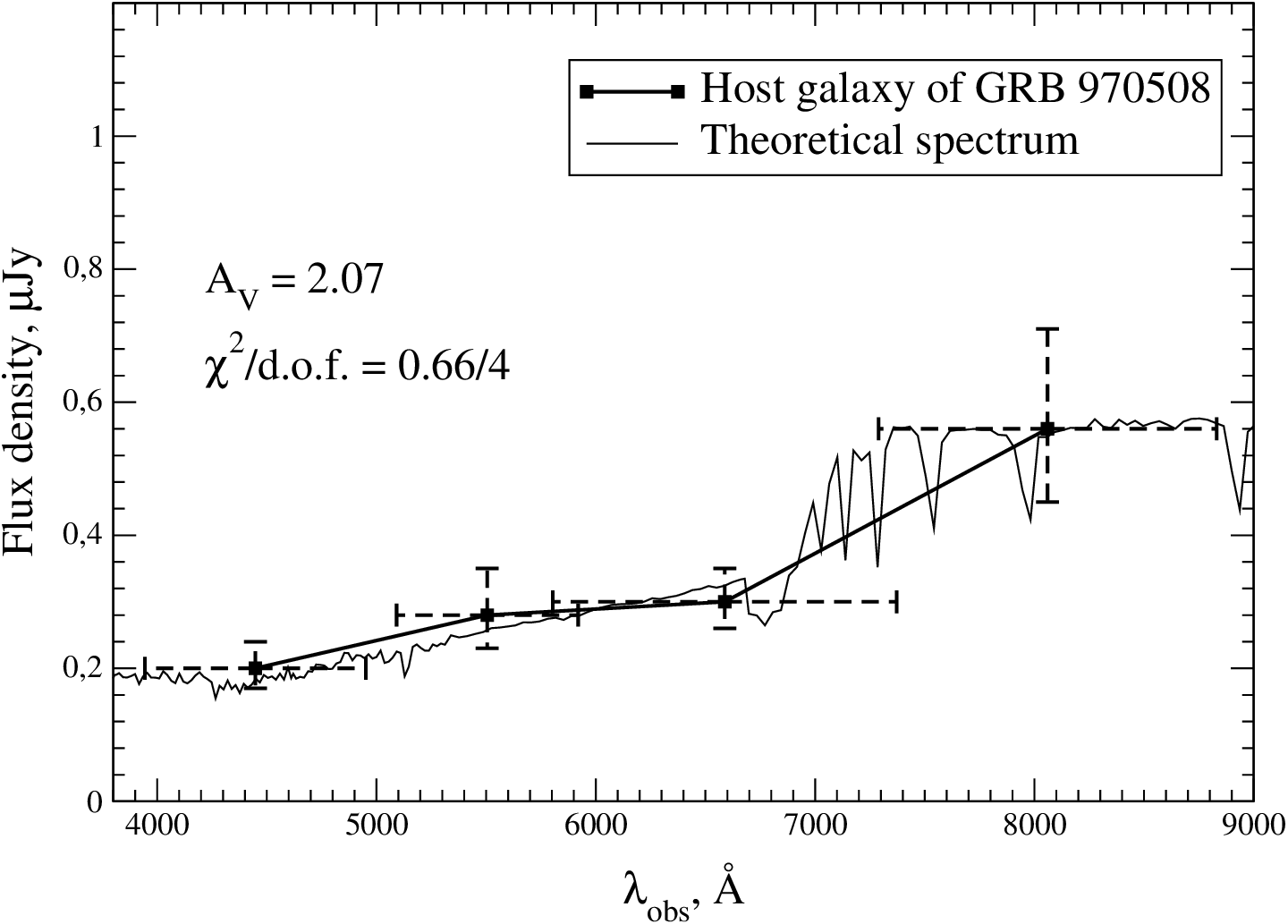}
\\
\parbox[t]{0.47\textwidth}{
\caption{A comparison of the GRB~970508 host galaxy broad-band
	 rest-frame ($z=0.835$) flux
	 spectrum with the SED of S5 template galaxies
	 (see \cite{Connoly}). Fluxes of S5 template were scaled for the
	 best fitting.
\label{grb970508_S5}}}
\hfill
\parbox[t]{0.47\textwidth}{
\caption{
The best fit for the model SED to the $BVRI$ photometry of
the host galaxy GRB~970508,
assuming the extinction law from \cite{Calz2000}. Wavelengths are in the
observed frame.\label{GRB970508_Model_Calz}}}
\end{figure*}

The multiband observations of GRB host galaxies were performed
with the 6-m telescope of SAO RAS in 1998--2000.
In details the data reduction and photometry of the host galaxies are described
in \cite{Sokolov2000}.
As {\it a first approximation} for an estimate of internal extinction,
we compared our broad-band flux spectrum of the host galaxy of
GRB~970508 (as an example) to SEDs of local starburst galaxies.
We used S1, S2, S3, S4, S5, S6 averaged template SEDs for the local
starburst galaxies from \cite{Calz1994}.
The spectra of local starburst were grouped according to increasing
values of the color excess $E(B-V)$.
Fig.~\ref{grb970508_S5} demonstrates the best fitting
(minimum of $\chi^2/d.o.f$) of our broad-band flux spectrum by the starburst
template galaxies (in Fig.~\ref{grb970508_S5} and \ref{GRB970508_Model_Calz}
FWHM of each band is shown, taking into account $z$, by dashed horizontal lines
with bars).
Using $E(B-V)$ for the S5 template and reddening laws from \cite{Cardelli}
and \cite{Calz2000} we can estimate the value of $A_V$, which is in ranges
$A_V = 1.58 \div 1.86$ and $A_V = 2.07\div 2.43$, respectively.
The same comparisons of the $BVRI$ broad-band spectra
with the local template SEDs allows us to derive K-correction and absolute
magnitudes.In Table~\ref{summary} we present the observed $R$-band
magnitudes and the estimates of $M_{B_{rest}}$ derived by us \cite{Sokolov2000}
and other authors for eight GRB host galaxies.

\begin{table*}
\caption{Observed and absolute magnitudes of GRB host galaxies }
\label{summary}
\centering
\begin{tabular}{lccl}
\hline
\hline
Host       &  observed magn.~~~ &  absolute magn.~~~ &  reference \\
	   &   $R$              &  $M_{B_{rest}}$    &            \\
\hline
GRB~970228 &  24.6$\pm$0.2      &  -18.6  & $R$: \cite{Galama}, $M_{B_{rest}}$: \cite{Bloom2000a} \\
GRB~970508 &  24.99$\pm$0.17    &  -18.6  & \cite{Sokolov2000} \\
GRB~991208 &  24.36$\pm$0.15    &  -18.8  & \cite{Sokolov2000} \\
GRB~990712 &  21.80$\pm$0.06    &  -19.9  & \cite{Hjorth2000}\\
GRB~980613 &  23.58$\pm$0.1     &  -20.8  & \cite{Sokolov2000} \\
GRB~990123 &  24.47$\pm$0.14    &  -20.9  & \cite{Sokolov2000} \\
GRB~971214 &  25.69$\pm$0.3     &  -21.1  & $R$: \cite{Sokolov2000}, $M_{B_{rest}}$: \cite{Kulkarni} \\
GRB~980703 &  22.30$\pm$0.08    &  -21.3  & \cite{Sokolov2000} \\
\hline
\hline
\end{tabular}
\end{table*}

As {\it a second approximation } for the estimate of internal extinction, we
constructed a set of model theoretical templates for the host galaxy of
GRB~970508 using population synthesis modeling.
We used the PEGASE package (\cite{PEGASE})
and the following assumptions: the Sun metalicity, instantaneous burst of star
formation, Salpeter initial mass function with the low and high mass cut-offs
to be 0.1\,\Myr
and 120\,\Myr, respectively; cosmology with $H_0$=60\,km\,s$^{-1}$\,Mpc$^{-1}$,
$\Omega_M$=0.3 and $\Omega_\Lambda$=0.7. For the calculations of the resulting
SED we applied a two-component model: the first component is just a burst
("burst" component) of star formation and the second one is an old
("old" component) stellar population. Both give corresponding contributions
into the resulting continuous SED.
The "burst" component is responsible for emission lines and nebular continuum.
For this reasons, we roughly fixed the "burst" component parameters using
the luminosity of the forbidden emission line [\ion{O}{ii}] by fitting to its
observed flux from \cite{Bloom98b} and taking into account the assumed
reddening laws.
With the constructed set of the theoretical templates we found the
minimum of $\chi^2/d.o.f$ for the $BVRI$ broad-band flux spectrum of the
GRB host in two ranges of $A_V$ obtained from the comparison with local
starburst templates in {\it a first approximation}:
$A_V=2.07\div 2.43$ and $A_V=1.58\div 1.86$ for two reddening laws
\cite{Calz2000} and \cite{Cardelli}.
According to our method, we derived the best fit parameters of the
model SEDs which are given in Table~\ref{model}.
In Fig.~\ref{GRB970508_Model_Calz} we plot the model SED in the case of the
reddening curve from \cite{Calz2000}. As it can be seen, the $BVRI$ broad-band
flux spectrum of the host galaxy of GRB~970508 is best fitted by the
theoretical template with sufficiently high internal extinction.
We notice that the best fit parameters correspond to the $A_V$, which
indeed, lies within the range of $A_V$ derived from the comparison with
the S5 template SED. Taking into account the reddening curve from
\cite{Calz2000}, $A_V$ from Table~\ref{model} and the lower limit of SFR from
\cite{Bloom98b}, we can estimate the extinction-corrected SFR
(star formation rate) as follows: SFR$_{corr}\approx$17\,\Myr.
\begin{table}[t]
\caption{The best fit model parameters}
\label{model}
\centering
\begin{tabular}{rcl}
\hline
\hline
reddening curve~~~                 & $A_V$~~~ & $\chi^2/d.o.f.$ \\
				   &          &  \\
\hline
Cardelli et al, \cite{Cardelli}~~~ & 1.6      & 1.03/4 \\
Calzetti et al, \cite{Calz2000}~~~ & 2.07     & 0.66/4 \\
\hline
\hline
\end{tabular}
\end{table}

We emphasize that only the simplest model assumptions
were made
and we did not include in the modeling other possibilities (e.g. exponentially
decreasing star formation scenario, subsolar metallicity). For comparison of
our value of $A_V$, we draw attention that in the case of the host
galaxy of GRB~990712 the extinction was obtained to be
$A_V = 3.4^{+2.4}_{-1.7}$ for the extinction law from \cite{Cardelli}
(see \cite{Vreeswijk}), which is about 2 times higher than our one
($A_V = 1.6$ for the same law). It should be noted in conclusion that
the comparison of $BVRI$ broad-band flux spectrum of the host galaxy of
GRB~970508 with local starburst templates and theoretical templates shows
that it is likely to be of great importance to take into account internal
extinction in GRB host galaxies.

{\it Acknowledgements:} This work was supported by INTAS N96-0315,
"Astronomy" Foundation (grant 97/1.2.6.4), RFBR N98-02-16542, RFBR N00-02-17689.

\end{document}